\documentclass{elsart}

\usepackage{graphicx,amssymb,amsmath}

\begin{document}
\begin{frontmatter}
\title{Impact of bistability in the synchronization \\
       of chaotic maps with delayed coupling} 
\author[1,2]{Pedro G.~Lind\corauthref{cor}},
\corauth[cor]{Corresponding author}
\ead{lind@icp.uni-stuttgart.de}
\author[1]{Ana Nunes},
\author[3]{Jason A.C.~Gallas}
\address[1]{Centro de F\'{\i}sica Te\'orica e Computacional,\\
            Av.~Prof.~Gama Pinto 2, 1649-003 Lisbon, Portugal,}
\address[2]{Institute for Computational Physics,\\
            Universit\"at Stuttgart, Pfaffenwaldring 27, 
            D-70569 Stuttgart, Germany,}
\address[3]{Instituto de F\'\i sica, Universidade Federal do 
            Rio Grande do Sul,\\ 91501-970 Porto Alegre, Brazil.}
\begin{abstract}
We investigate the impact of bistability in the emergence of synchronization
in networks of chaotic maps with delayed coupling.
The existence of a single finite attractor of the uncoupled map is found
to be responsible for the emergence of synchronization.
No synchronization is observed when the local dynamics 
has two competing chaotic attractors whose orbits are dense
on the same interval. 
This result is robust for regular networks with variable
ranges of interaction and for more complex topologies.
\end{abstract}
\end{frontmatter}


Synchronization of chaotic systems  has been frequently
studied nowadays because of its relevance in many situations, 
e.g.~in nonlinear optics and fluid dynamics \cite{boccaletti02}. 
Since a realistic coupling between chaotic systems should consider 
finite-time 
propagation of the interaction, particular attention has 
been recently given to time-delayed coupling~\cite{atay04,masoller05}.
For instance, delayed coupling was used to study the emergence of 
multistability in noisy bistable elements~\cite{huber03} and to analyze 
anticipating synchronization of two excitable systems~\cite{ciszak04}.
It is already known that for continuous systems the stability conditions 
of coupled elements is not influenced by the time-delay of the coupling 
when the elements are randomly coupled~\cite{jirsa04}, and 
that synchronization is independent of the topology but depends
on the average number of neighbors~\cite{masoller05}.
For discrete systems modeled by coupled map lattices, it was found that 
even with random delays, for adequate coupling strength an array of chaotic 
logistic maps is able to synchronize~\cite{masoller05}. 

Here, we present robust numerical evidence that 
synchronization effects observed in lattices of coupled logistic
maps are strongly dependent of the local  map.
Specifically, we show that synchronization may be easily destroyed
by simply substituting the logistic map by a bistable map with
two competing chaotic attractors.
This is true both for  regular and for complex networks.

Our model consists of the usual array of coupled chaotic maps defined by
\begin{equation}
x_{t+1}(i) = (1-\varepsilon) f(x_t(i)) +
             \frac{\varepsilon}{\sum_{j=1}^N \eta_{ij}}
             \sum_{j\neq i} \eta_{ij} \omega_{ij} f(x_{t-\tau}(j)),
\label{model}
\end{equation}
where $\tau$ represents the delay,
$f(x_t)$ denotes the local map, $\varepsilon$ is 
the coupling strength and $N$ is the total number of nodes.
The adjacency matrix $\{ \eta_{ij} \}$ and the coupling strength
$\omega_{ij}$ between nodes $i$ and $j$ depend on the network topology
and for all $i$ one has $\sum_{j=1}^N \omega_{ij}=N-1$.
As usual, synchronized solutions are detected when the standard deviation 
of the amplitudes $x_t$ is numerically zero, i.e.~when
$\sigma^2(t) = \tfrac{1}{N-1}\sum_j (x_t(j)-\langle x_t \rangle )^2 
\lesssim 10^{-20}$. 
\begin{figure}[t]
\begin{center}
\includegraphics*[width=6.0cm]{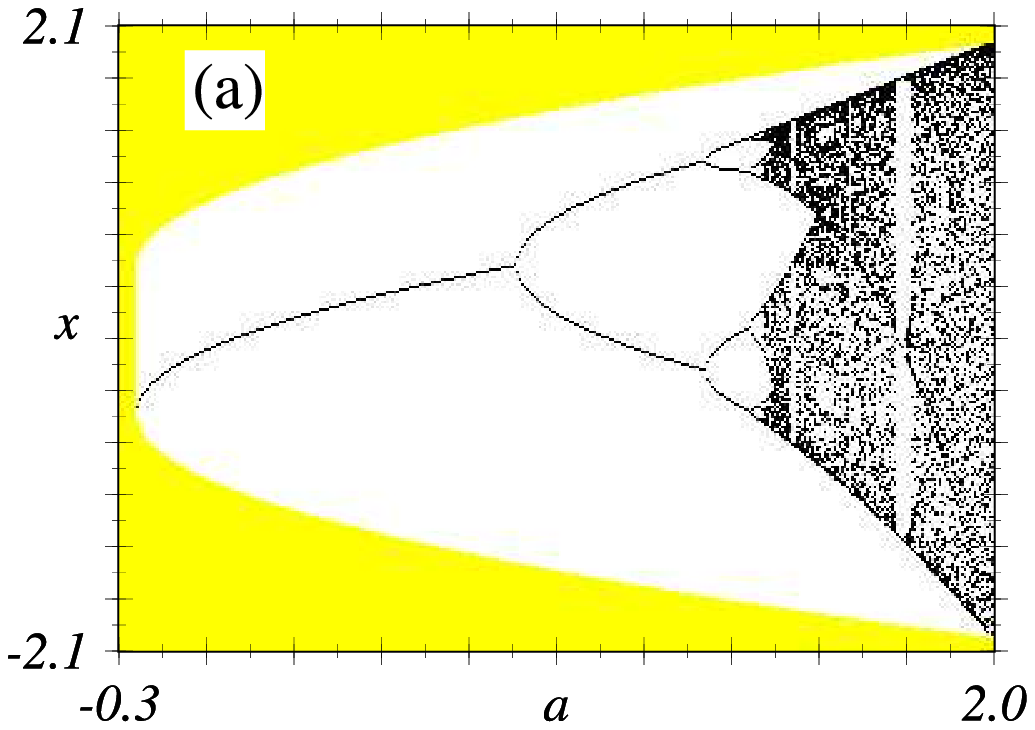}%
\includegraphics*[width=6.0cm]{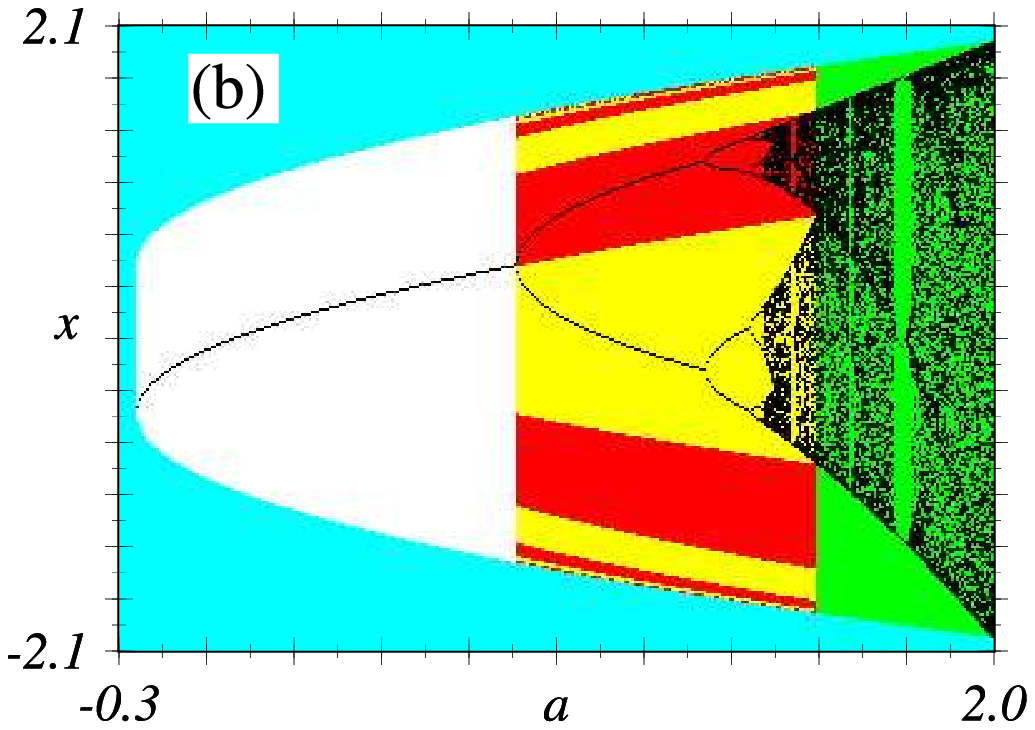}
\end{center}
\caption{\protect Bifurcation diagrams of 
         {\bf (a)} the logistic map $x_{t+1}=a-x_t^2$
         with one single finite stable attractor, and of
         {\bf (b)} the quartic map $x_{t+1}=a-(a-x_t^2)^2$, 
         where for $a>0.75$ two stable attractors co-exist.
         Different basins of attraction are represented
         with different tonalities.}
\label{fig1}
\end{figure}
\begin{figure}
\begin{center}
\includegraphics*[width=4.2cm]{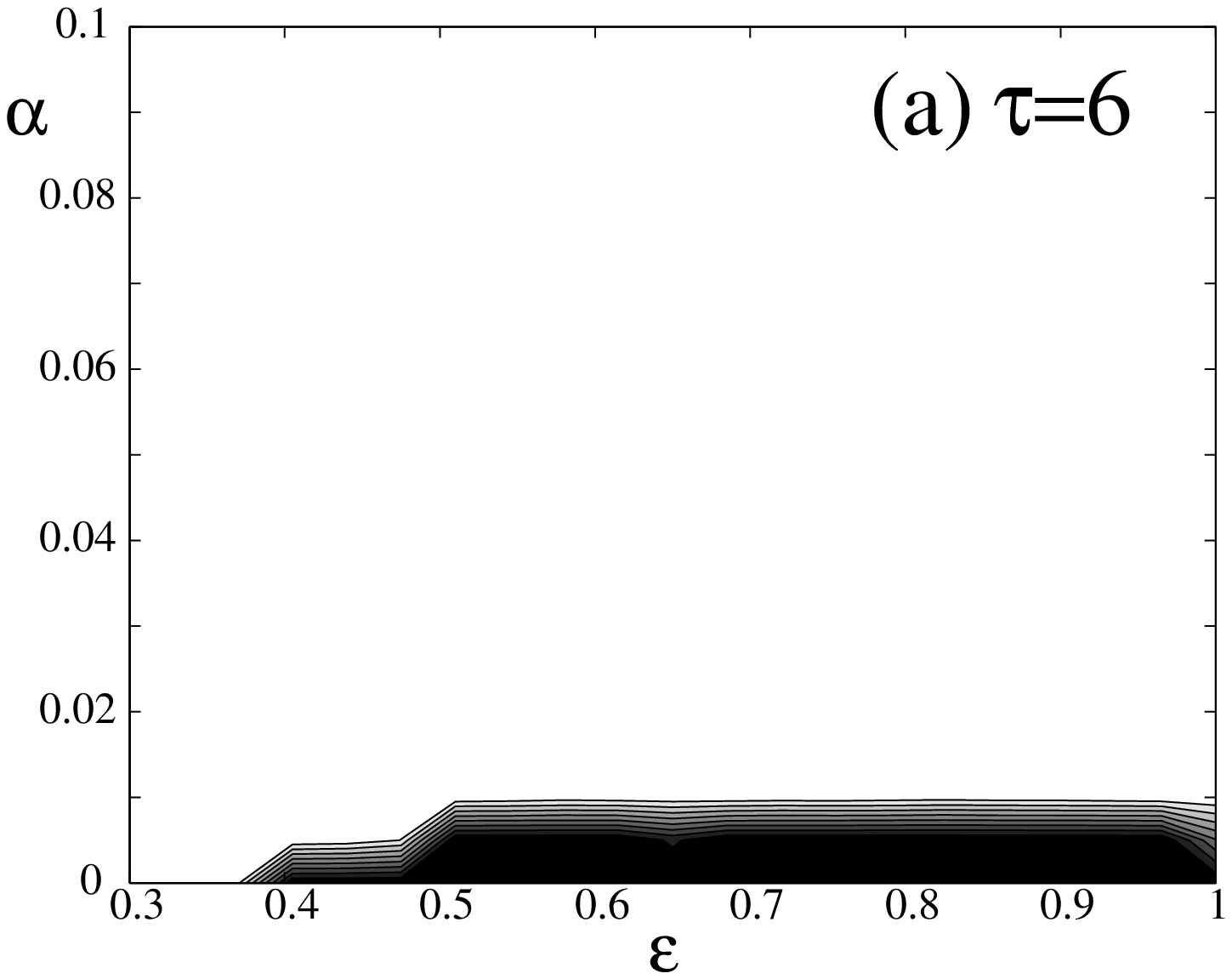}%
\includegraphics*[width=4.2cm]{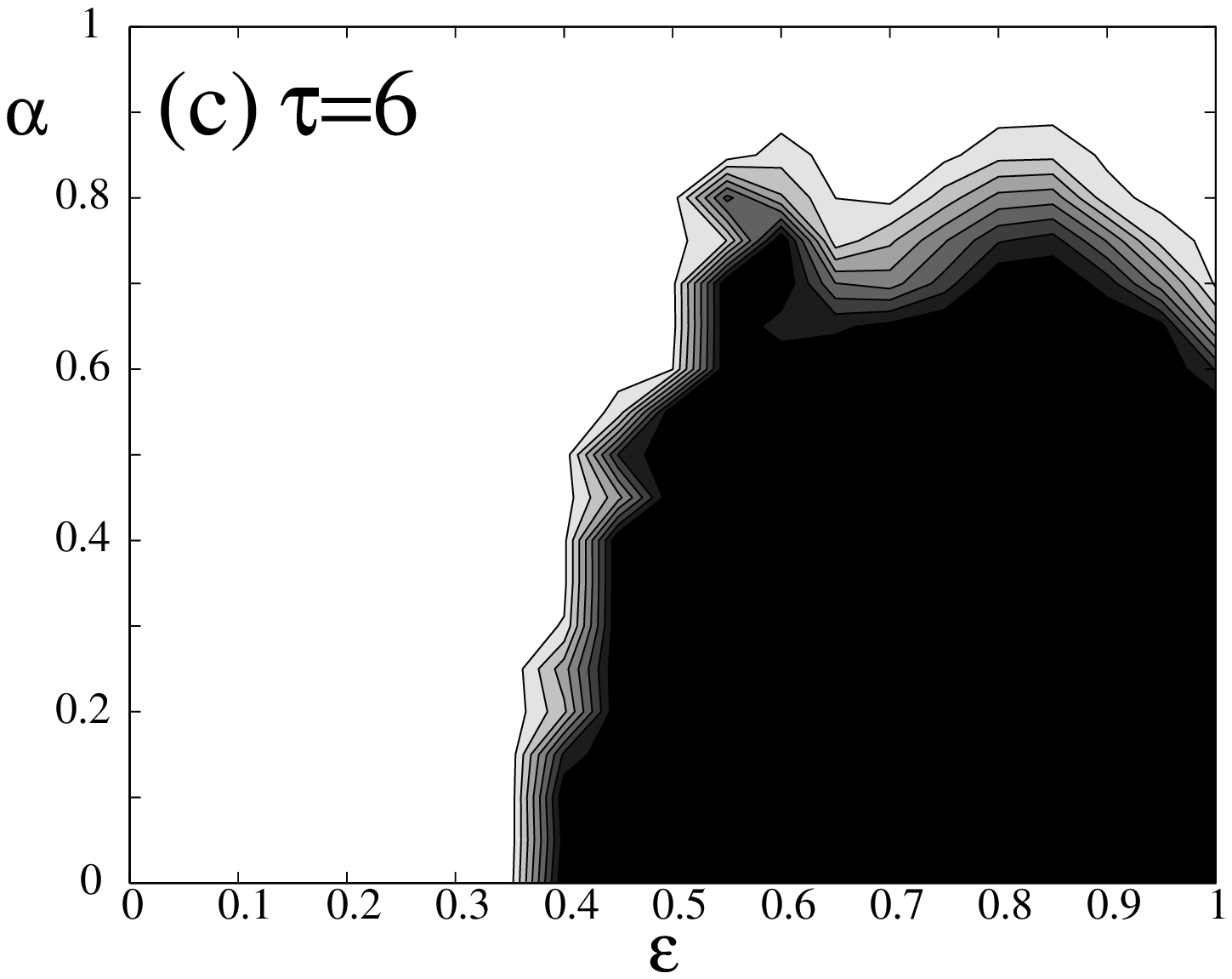}%
\includegraphics*[width=4.2cm]{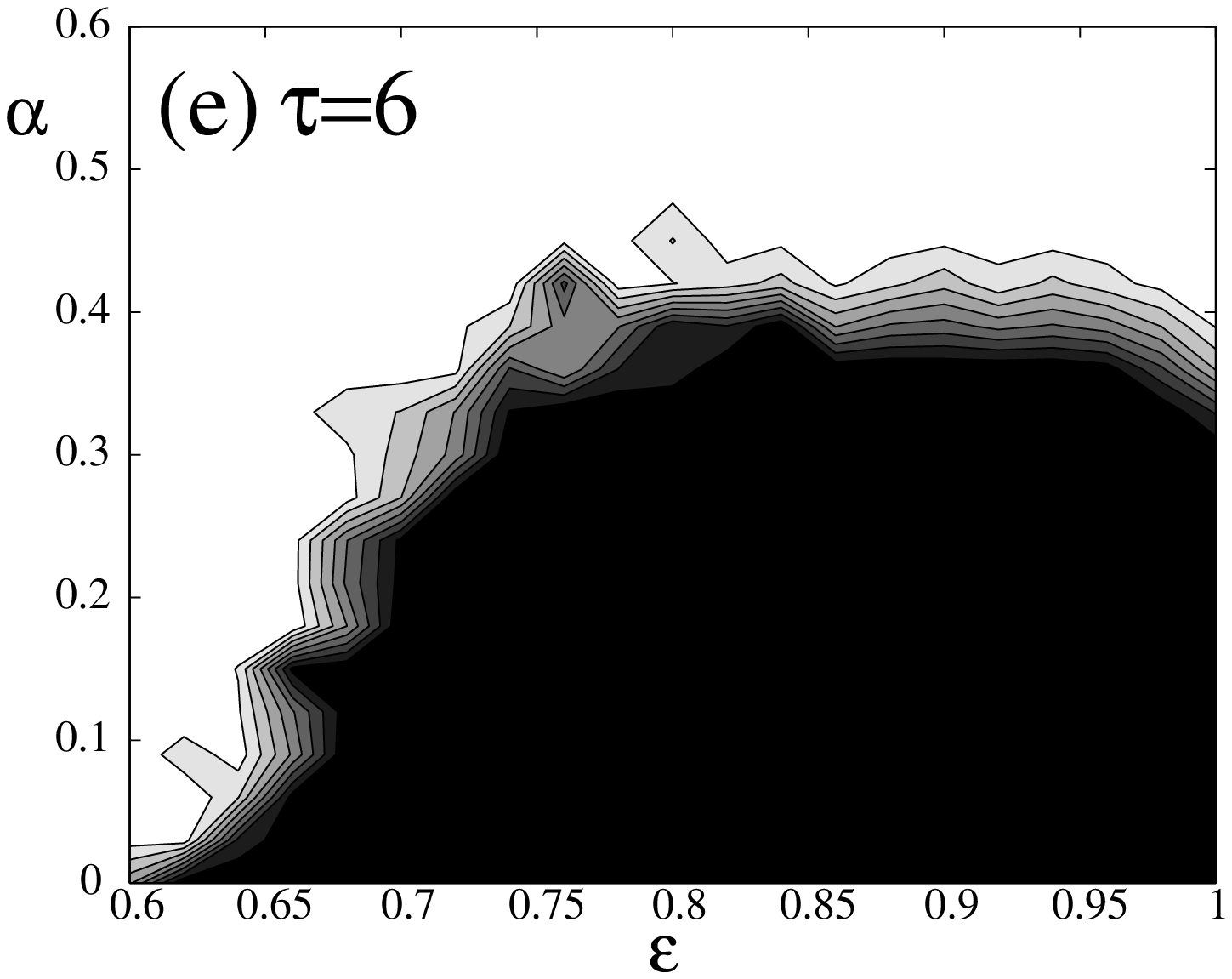}
\includegraphics*[width=4.2cm]{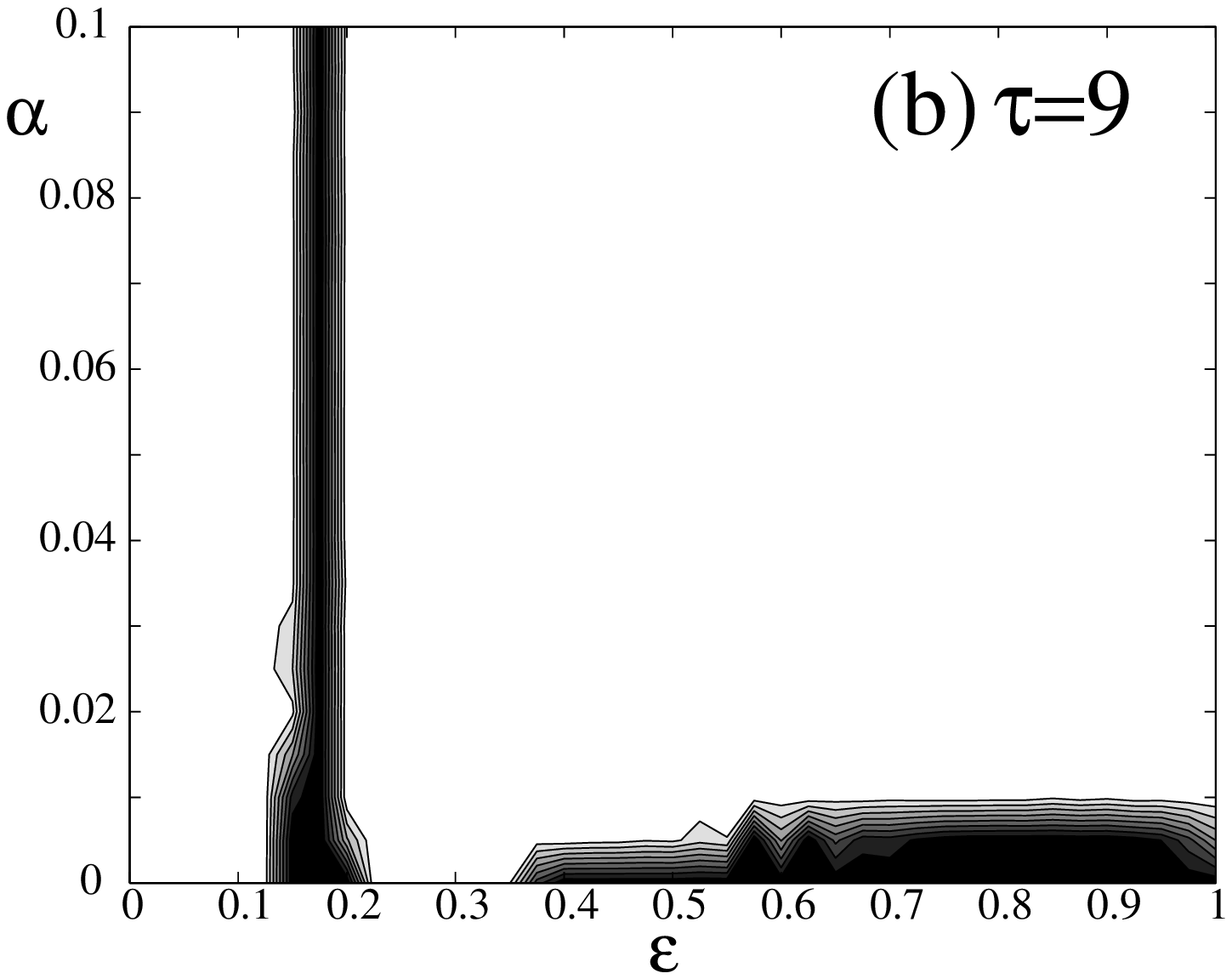}%
\includegraphics*[width=4.2cm]{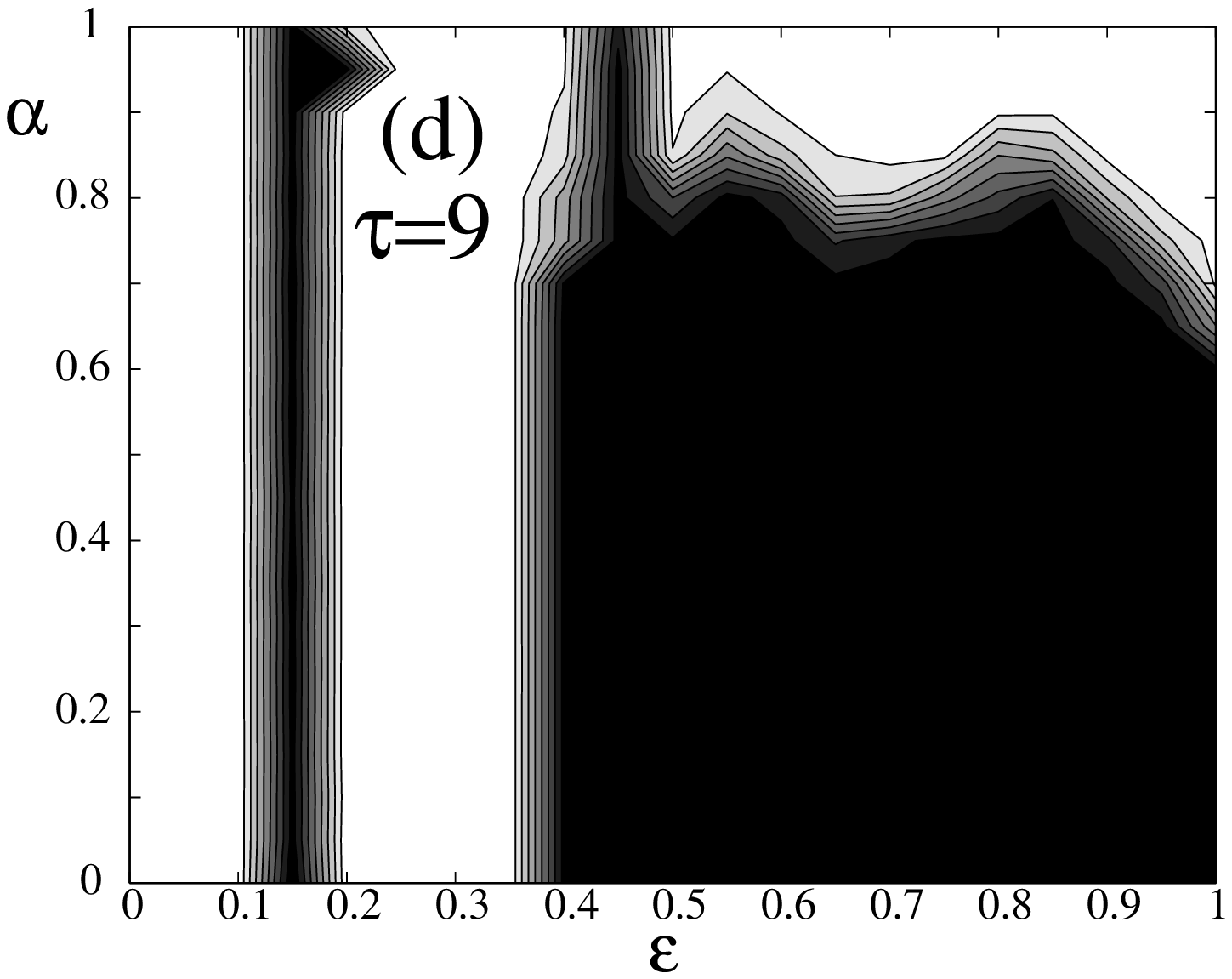}%
\includegraphics*[width=4.2cm]{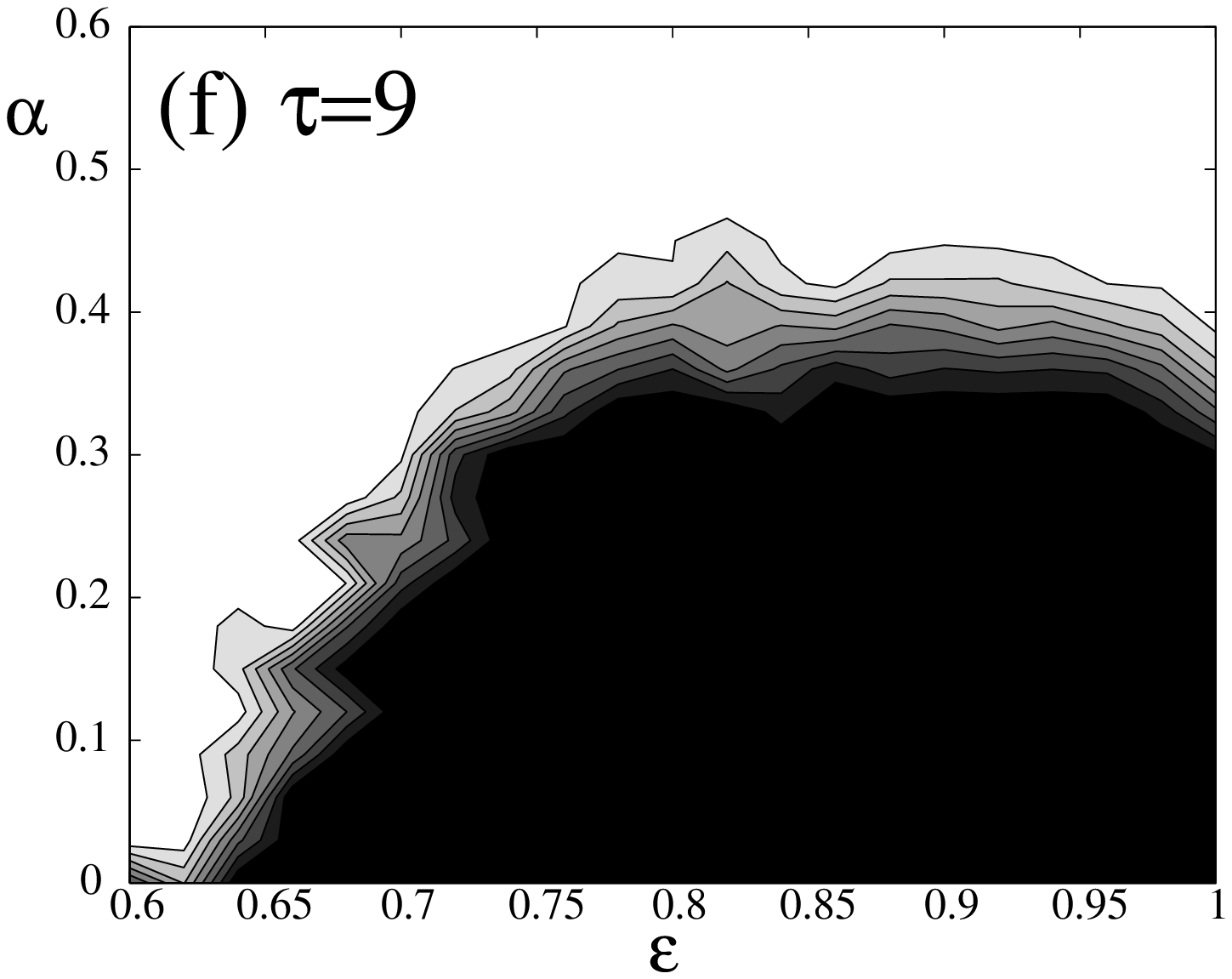}
\end{center}
\caption{\protect Regions of full-synchronization (black) for regular
         networks with 
         {\bf (a-b)} exponential coupling using logistic maps,
         {\bf (c-d)} polynomial coupling with logistic maps and
         {\bf (e-f)} polynomial coupling with quartic maps.
         Here $a=2$, $N=10^3$, transients are $10^4$, samples 
         contain $50$ initial configurations. $\tau$ is the delay
         defined in Eq.~(\ref{model}).}
\label{fig2}
\end{figure}

In this framework,
we contrast the properties of the standard logistic map $f_l(x_t)=a-x_t^2$ 
characterized by a single finite attractor,
with those found for the quartic map $f_q(x_t)=a-(a-x_t^2)^2$ 
having  two stable finite  attractors (bistability).
Although the bifurcation diagram of these maps look numerically the same
(see Fig.~\ref{fig1}), at $a=0.75$ the logistic map suffers a 
period-doubling bifurcation while the quartic map shows a tangent  bifurcation
where the fixed point splits into a {\it pair\/} of fixed points,
each one with its own basin of attraction and doubling cascade. 
By contrast with the logistic map, the 
quartic map has two stable finite attractors beyond $a>0.75$.
In particular, in the fully chaotic regime ($a=2$) the two chaotic orbits of
the quartic map are dense in the interval $[-2,2]$, i.e.~any subinterval
intersects both basins of attraction.
Therefore, one could expect that for $a=2$ the two maps are 
numerically indistinguishable when introduced in Eq.~(\ref{model}). 
However, as we show below this is not the case.

We start by considering a regular network  where each node $i$ is
symmetrically coupled to its neighbors, labeled as $j$, 
in both short- and long-range interaction regimes.
For short-range coupling, the coupling strength decreases with the distance 
$\vert i-j \vert_{\hbox{mod}N/2}$, 
i.e.~$\omega_{ij}\propto\exp{(-\alpha\vert i-j \vert)}$, while for long-range 
coupling the decrease is polynomial, 
$\omega_{ij}\propto\vert i-j \vert ^{-\alpha}$.
In both cases, 
$\alpha$ controls the range of interaction: $\alpha=0$
yields the fully and uniformly connected network while for $\alpha\to\infty$
maps are decoupled.
\begin{figure}
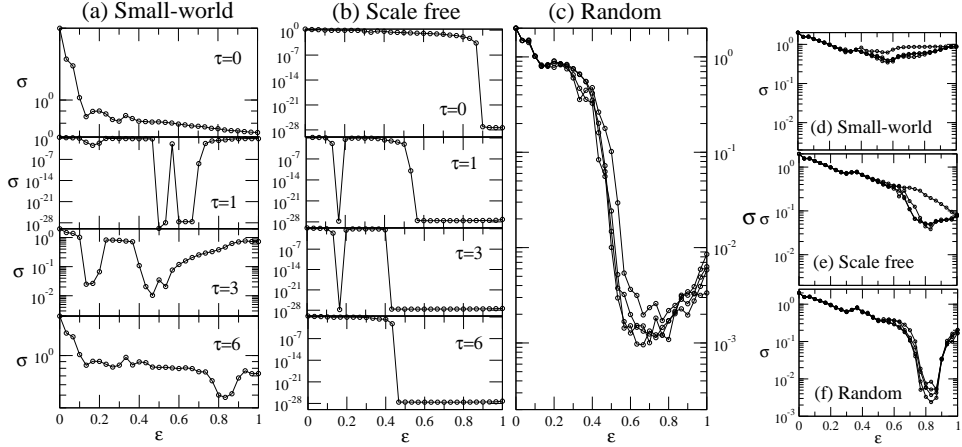

\begin{center}
\includegraphics*[width=10.0cm]{fig3a_delay_lng.eps}%
\includegraphics*[width=2.7cm]{fig3b_delay_lng.eps}
\end{center}
\caption{\protect 
         Standard deviation $\sigma$ of the node amplitudes as a function
         of the coupling strength in complex networks using
         {\bf (a-c)} the logistic map and
         {\bf (d-f)} the quartic map.
         In all cases $\langle k\rangle=10$ and the conditions of 
         Fig.~\ref{fig2} were used.
         The small-world network is constructed using the algorithm in
         \cite{newman} with a probability $p=0.05$ for acquiring random 
         connections, while the scale-free network is constructed from 
         the Albert-Barab\'asi model.}
\label{fig3}
\end{figure}
\begin{figure}
\begin{center}
\includegraphics*[width=3.5cm]{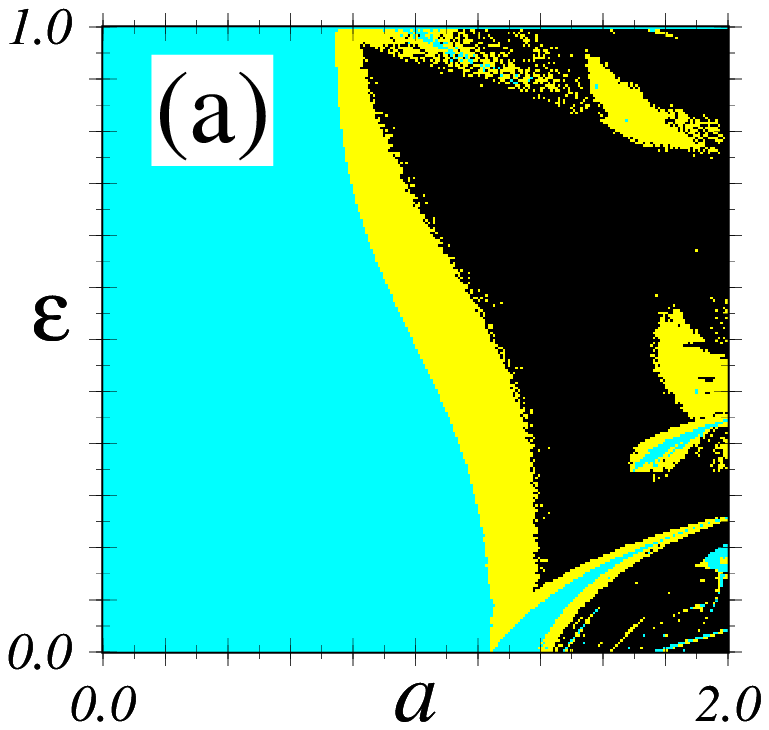}%
\includegraphics*[width=3.5cm]{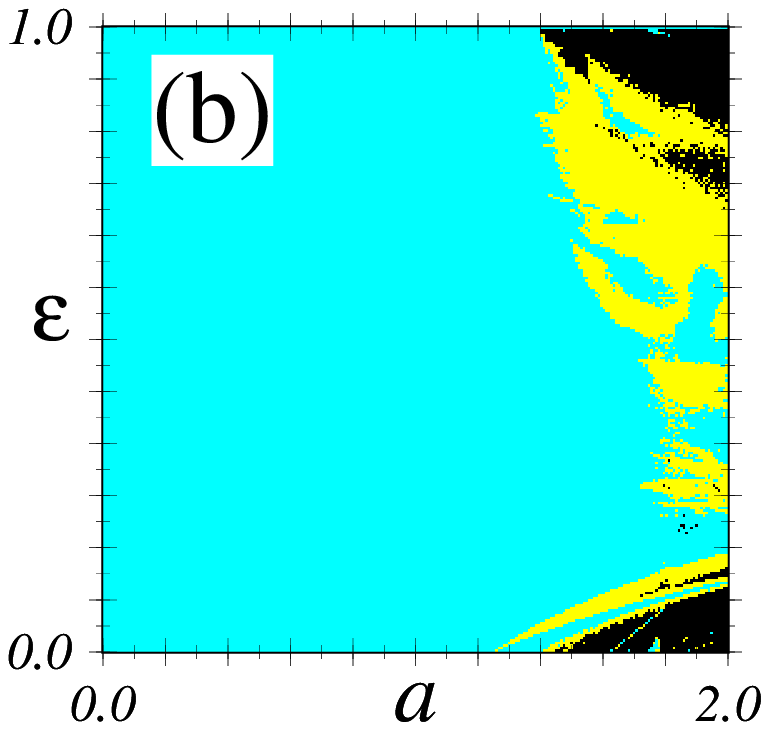}%
\includegraphics*[width=3.5cm]{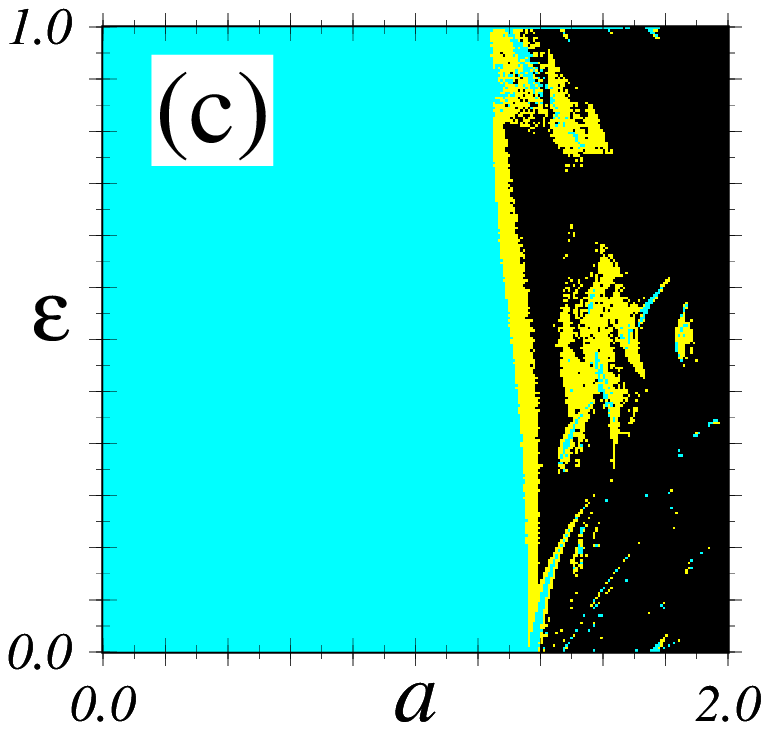}%
\includegraphics*[width=3.5cm]{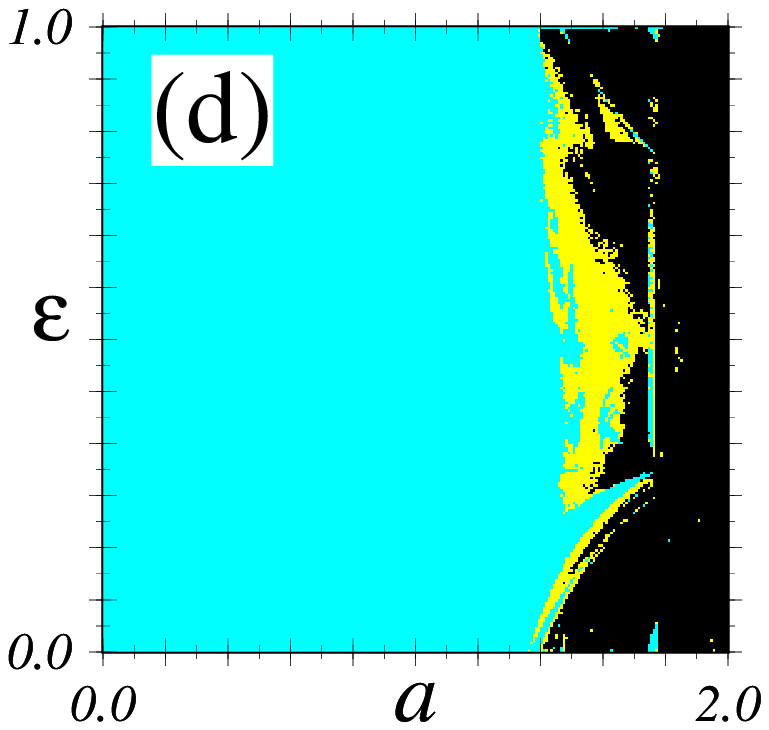}
\end{center}
\caption{\protect 
         Illustration of periodic (gray), chaotic (white)
         and hyperchaotic (black) regions of the synchronized 
         solutions in parameter space ($a$ and $\varepsilon$). 
         Logistic map:
         {\bf (a)} $\tau=1$ and
         {\bf (b)} $\tau=2$.
         Quartic map:
         {\bf (c)} $\tau=1$ and
         {\bf (d)} $\tau=2$.
         For each pair of values, Lyapunov exponents were computed from 
         intervals of $10^3$ time-steps after discarding transients of 
         $10^3$ time-steps.}
\label{fig4}
\end{figure}

Full synchronization is observed in the black regions in Fig.~\ref{fig2}.
Exponential coupling using logistic maps is illustrated in 
Figs.~\ref{fig2}(a-b),
while 
polynomial coupling using logistic and quartic maps is shown in
Figs.~\ref{fig2}(c-d) and (e-f)  respectively. 
For quartic maps with exponential coupling the results are independent of 
$\tau$ and are similar to those of Fig.~\ref{fig2}(a).
For other even and odd values of $\tau$,  one obtains plots similar to 
the ones for $\tau=6$ and $\tau=9$ respectively.
From Fig.~\ref{fig2} one clearly sees that, while for exponential coupling
synchronization is observed only in a neighborhood of $\alpha=0$, for 
polynomial coupling, the synchronization region spreads in a wider
range of $\alpha$-values, since the coupling strength decreases more slowly 
with the distance. 
Moreover, there is different behavior for the two
types of local dynamics: 
for the logistic map there is a synchronization region within 
$0.1\lesssim\varepsilon\lesssim 0.2$ while for the quartic map there is not.
In the particular case of nearest-neighbors coupling this is almost
the unique synchronization region corroborating recent findings \cite{atay04}.

For complex topologies bistability also suppresses the emergence of
synchronized solutions as shown in Fig.~\ref{fig3}.
In order to ascertain the robustness of our results, we consider three 
different types of complex topologies, namely a random topology, a 
small-world topology and a scale-free network.
Figures \ref{fig3}(a-c) show the results for the logistic map, while
Figs.~\ref{fig3}(d-f) show the results for the quartic map.
In all cases of Fig.~\ref{fig3} the average degree is fixed at
$\langle k\rangle=10$ neighbors, and we have taken,
for simplicity, $\omega_{ij}=1-\delta_{ij}$ (uniform coupling).

In addition to suppression of synchronization in the case of the
quartic map, one observes, for the particular case of scale-free 
networks of logistic maps, a very abrupt transition to synchronization.
This transition, already reported in previous studies~\cite{hans},
is still not fully understood. Our simulations indicate that
the coupling strength threshold is independent of the network size.


For all the cases above, the synchronized solution of Eq.~(\ref{model}) 
evolves according to $X_{t+1}=(1-\varepsilon)f(X_t)+\varepsilon f(X_{t-\tau})$,
involving the two variables $X_t$ and $X_{t-\tau}$.
Therefore, the corresponding Jacobian yields two different Lyapunov
exponents: if both exponents are negative the orbit is periodic, if
one of the exponents is positive the orbit is chaotic and when both
are positive the orbit is hyperchaotic.
Figures \ref{fig4}(a-b) show where each of the three regimes occurs for the 
logistic map either with even and odd time delays while 
Figs.~\ref{fig4}(c-d) illustrate the same situations for the quartic map.
Clearly, for $a=2$ the quartic map shows hyperchaos in all the range of 
$\varepsilon$ values, while in the logistic case only certain $\varepsilon$
ranges exhibit hyperchaos. This fact could be related with the significant
suppression of synchronized solutions when bistability sets in.
For other odd or even time delays similar plots are obtained.

In conclusion, we found that bistable local dynamics in networks of
chaotic maps with delayed coupling destroys synchronization 
for a wide variety of situations.
In the literature, synchronization in networks with delayed 
coupling was only studied so far under the very particular unimodal 
local dynamics.
As shown here, however, care should be taken when interpreting results
obtained with logistic maps because synchronized solutions are too
dependent of  the maps  ruling the dynamics.

The authors thank M.T.~da Gama for useful discussions. 
PGL thanks Funda\c{c}\~ao para a Ci\^encia e a Tecnologia (FCT), Portugal, 
for a postdoctoral fellowship.
JACG thanks Conselho Nacional de Desenvolvimento Cient\'{\i}fico e 
Tecnol\'ogico, Brazil, for a Senior Research Fellowship.



\begin{thebibliography}{00}

\bibitem{boccaletti02} S.~Boccaletti, J.~Kurths, G.~Osipov, 
                       D.L.~Valladares and C.S.~Zhou,
                       Phys.~Rep.~366 (2002) 1.

\bibitem{atay04} F.M.~Atay and J.~Jost,
                 Phys.~Rev.~Lett.~92 (2004) 144101.

\bibitem{masoller05} C.~Masoller and A.C.~Mart\'{\i},
                     Phys.~Rev.~Lett.~94 (2005) 134102.


\bibitem{huber03} D.~Huber and L.S.~Tsimring,
                  Phys.~Rev.~Lett.~91 (2003) 260601.

\bibitem{ciszak04} M.~Ciszak, F.~Marino, R.~Toral and S.~Balle,
                   Phys.~Rev.~Lett.~93 (2004) 114102.

\bibitem{jirsa04} V.K.~Jirsa and M.~Ding,
                  Phys.~Rev.~Lett.~93 (2004) 070602.


\bibitem{newman} M.E.J.~Newman and D.J.~Watts,
                 Phys.~Rev.~E 60, 7332 (1999). 


\bibitem{hans} P.G.~Lind, J.A.C.~Gallas and H.J.~Herrmann,
               Phys.~Rev.~E 70, 056207 (2004).

\end{thebibliography}
\end{document}